\def\be{\begin{equation}}
\def\ee{\end{equation}}
\def\figloc#1#2{\epsfysize=5in
    \centerline{\epsfbox{fig#1.ps}}
    \centerline{Figure #1}
    {\raggedright\it   #2 }
    \bigskip
    }

\documentstyle[preprint,aps, epsf]{revtex}

\begin{document}
\title{ Choptuik scaling in null coordinates}

\author{David Garfinkle}
\address{Dept. of Physics, Oakland University, Rochetster, MI 48309 \\
email: garfinkl@vela.acs.oakland.edu}

\maketitle

\bigskip
\begin{abstract}
A numerical simulation is performed of the gravitational collapse of a
spherically symmetric scalar field.  The algorithm uses the null initial
value formulation of the Einstein-scalar equations, but does {\it not} use
adaptive mesh refinement.  A study is made of the critical phenomena found
by Choptuik in this system.  In particular it is verified that the critical
solution exhibits periodic self-similarity.  This work thus provides a
simple algorithm that gives verification of the Choptuik results.
\end{abstract}

\section{Introduction}

Recently Choptuik has discovered scaling behavior in the collapse of a
spherically symmetric scalar field to form a black hole.\cite{Chop}  Choptuik
numerically evolves a family of initial data parametrized by
$ p $ to find that the mass of the black hole
is
$ M \propto {{(p - {p^*} )}^\gamma} $
where
$ p^* $
is the critical value of
$ p $
and
$ \gamma $
is the scaling exponent.  For the data with
$ p = {p^*} $,
a zero mass singularity forms.  This critical solution has the
property of periodic self-similarity: the scalar field evolves, after
a certain amount of time, to a copy of its profile with the scale of space
shrunk.  Similar results have been found by Abrahams and Evans\cite{Evans}
for vacuum axisymmetric gravitational collapse.

To treat the critical solution the parameter
$ p $
must be tuned to
$ p^* $
to great accuracy.  In addition the size of features must be resolved
on extremely small scales.  Therefore one might worry that the periodic
self-similarity of the critical solution
could be an artifact of the numerical algorithms used rather than the actual
behavior of the
collapse of a scalar field.  The most srtaightforward way to show that the
results of \cite{Chop} are not numerical artifacts is to perform a numerical
treatment of the same physical problem using a completely different algorithm.
However, any accurate treatment of the critical solution must be able to
resolve features on extremely small scales.  Choptuik achieved this by using
an adaptive mesh refinement algorithm.  Since adaptive mesh
refinement algorithms are fairly
complicated it is not a trivial task to produce another adaptive
mesh refinement code
to redo the Choptuik result.  Even with such a code one might worry that the
results are an artifact of adaptive mesh refinement.  In this paper I
present results
of a numerical simulation of the critical collapse of a scalar field.  The
algorithm uses the null initial value formulation of the problem, rather than
the spacelike initial value formulation used in \cite{Chop}.  In addition, no
adaptive mesh refinement is used.  I find results in agreement
with those of Choptuik.
Section II describes the null initial formulation of the collapse of a
spherically symmetric scalar field.  Section III is a description of the
numerical algorithm used to evolve this system.  Section IV contains the
results.

\section {Null initial value formulation}

The null initial value formulation for the collapse of a spherically
symmetric scalar field was worked out by
Christodoulou.\cite{DC}  In this section we review this formulation and
introduce the notation to be used in the rest of the paper.
The Einstein-scalar equations are
\be
{R_{ab}} = 8 \pi {\nabla _a} \Phi {\nabla _b} \Phi
\ee
where
$ R_{ab} $
is the Ricci tensor and
$ \Phi $
is the scalar field.

Null coordinates
$ u $
and
$ v $
are defined as follows:
$ u $
is the proper time of an observer at the origin and is constant on
outgoing light rays;
$ u = 0 $
is the initial data surface.  The coordinate
$ v $
is constant on ingoing light rays and is equal to
the usual area coordinate
$ r $
on the initial data surface.  (In what follows
$ r $
will be regarded as a function of
$ u $
and
$ v $).

For any quantity
$ f $
define
$ {\dot f} , \, {f \,  ' } $
and
$ \bar f $
by
\be
{\dot f} \equiv  {{\partial f} \over {\partial u}} \; \; \; ,
\ee
\be
{f '} \equiv  {{\partial f} \over {\partial v}} \; \; \; ,
\ee
\be
{\bar f} \equiv {1 \over r} {\int _0 ^v} f ( u, {\tilde v}) \, r '
(u, {\tilde v}) \, d {\tilde v} \; \; \; .
\ee
Let
$ h $
be the scalar such that
$ {\bar h} = \Phi $.
Then as a consequence of Einstein's equations the metric can be
written in terms of
$ h $
and
$ r $.
Define the quantities
$ q $
and
$ g $
by
\be
q \equiv {r ^{ - 1}} \, {{\left ( h \, - \, {\bar h} \right ) }^2} \; \; \; ,
\ee
\be
g \equiv \exp \left ( 4 \pi r {\bar q} \right ) \; \; \; .
\ee
Then the metric is
\be
d {s^2} = - \, 2 \, g \, {r \, '} \, d u \, d v \; + \; {r^2} \,  d
{\Omega ^2} \; \; \; .
\ee
Here
$ d {\Omega ^2} $
is the two-sphere metric.
Einstein's equations also provide evolution equations for
$ h $
and
$ r $
These are\cite{DC}
\be
{\dot h} = {1 \over {2 r}} \; \left ( g \, - \, {\bar g} \right ) \,
\left ( h \, - \, {\bar h} \right ) \; \; \; ,
\ee
\be
{\dot r} = - \; {1 \over 2} \; {\bar g} \; \; \; .
\ee

\section {Numerical methods}

A numerical simulation of these equations was first performed by Goldwirth and
Piran.\cite{GP}  A version of this algorithm was applied to
the Choptuik problem by
Grundlach, Price and Pullin.\cite{GPP}  However, the methods of
references\cite{GP,GPP} are not accurate enough to treat the critical
solution.  I will start by discussing those features that my algorithm has
in common with these earlier treatments and then present the new features that
give the improvements in accuracy.

Initial data for the Einstein-scalar equations is just the value of
$ h $
on the initial data surface.  One evolves these equations as
follows:\cite{GP,GPP} first one finds in succession the quantities
$ {\bar h} , \, q , \, g $
and
$ \bar g $.
This is done by evaluating the integrals for these quantities using Simpson's
rule for unequally spaced points.  Then the equations for
$ \dot h $
and
$ \dot r $
are used to evolve
$ h $
and
$ r $
forward one time step.  This process is iterated as many times as
necessary: {\it i.e.} until either a black hole forms or the field
disperses.  To find
whether a black hole forms one looks for a marginally outer trapped
surface. For spherical symmetry this is a surface for which
$ {\nabla ^a} r {\nabla _a} r = 0 $. (In practice the code cannot evolve up
to the marginally outer trapped surface so one looks for the condition
$ {\nabla ^a} r {\nabla _a} r \to 0 $.)
Since such a surface has
$ r = 2 M $
the mass of the black hole is then half the radius of the marginally
outer trapped surface.

In\cite{GPP} this method was used to study the scaling behavior of
the mass of the black hole.  However, the method is not accurate enough
for a treatment of
the critical solution.  I will now describe the sources of inaccuracy and
the methods that my code uses to overcome them.  One source of inaccuracy comes
from the fact that one divides by
$ r $
in evaluating the quantities
$ {\bar h} , \, q  $
and
$ \bar g $.
This leads to inaccuracies near
$ r = 0 $.
The second source of inaccuracy comes from the behavior of the
critical solution.  As the critical solution evolves its structure appears
on ever smaller spatial scales.  With a fixed spatial resolution there
comes a time when the number of grid points is not sufficient to resolve
the structure of the scalar field.

My code overcomes the first source of inaccuracy as follows: first expand
$ h $
in a Taylor series in
$ r $.
\be
h = {h_0} \; + \; {h_1} \, r \; + \; O({r^2} ) \; \; \; ,
\ee
Then the Taylor series for
$ {\bar h}, \, q , \, g $
and
$ \bar g $
are given by
\be
{\bar h}  = {h_0} \; + \; {1 \over 2} \; {h_1} \, r \; + \;
O ({r^2} ) \; \; \; ,
\ee
\be
q  = {1 \over 4} \; {h_1 ^2} \, r \; + \; O ({r^2} ) \; \; \; ,
\ee
\be
g  = 1 \; + \; {\pi \over 2} \; {h_1 ^2} \, r \; + \; O ({r^2} ) \; \; \; .
\ee
\be
{\bar g}  = 1 \; + \; {\pi \over 4} \; {h_1 ^2} \, r \; + \;
O ({r^2} ) \; \; \; .
\ee
Thus to find the behavior of all quantities near the origin one needs
to find only
$ h_ 0 $
and
$ h_1$.
This is done by fitting the first four values of
$ h $
to a line.  Equations (11-14) are then used to find the values of
$ {\bar h}, \, q , \, g $
and
$ {\bar g} \, $
for the first three values of
$ r $.
The Simpson's rule integration is then used for all other values of
$ r $.

The inaccuracy due to poor spatial resolution is solved as follows:  first note
that the inaccuracy is due to the size of the spatial structure being much
smaller than the overall size of the grid.  However, in the null initial value
formulation the grid points are tied to ingoing light rays.  Thus as the
system evolves the overall size of the grid becomes smaller and the
resolution improves.  The critical solution forms a zero mass singularity.
Consider the ingoing light ray that just barely hits this singularity.  Choose
the outermost gridpoint to correspond to this light ray.  Then the size of
the grid shrinks as much as the size of the spatial features and thus
the resolution is always good.  In practice one does not know beforehand
the position on the initial data surface of the light ray that just barely
hits the singularity.  Therefore I have run the code as follows:
First I make an estimate of where the outermost gridpoint should be and choose
the point to be slightly higher than the estimate.  This ensures that the
evolution will proceed for a while with good resoltion.  However, after a
certain amount of time, because the outermost grid point is too far, the
spatial features will come to occupy a relatively small number of grid points.
Observation of the size of the spatial features at this time allows a refined
estimate for the position of the outermost grid point.  This refined estimate
is then used to choose a (slightly too large) outermost gridpoint.  This
whole process is then iterated (a few times) until the outermost gridpoint
corresponds, with good accuracy to the light ray that just barely hits the
singularity.

In this way the overall size of the grid shrinks in proportion to the size of
the spatial features.  However, it is still the case that the number of
grid points decreases during the evolution.  This is because each
grid point corresponds to an ingoing light ray.  The grid point is
therefore lost when the light ray hits the origin.  This difficulty is
dealt with in the code as follows:  the evolution proceeds until half of
the grid points are lost.  These grid points are then interpolated halfway
in between each of the remaining grid points.  Thus over time the number of
gridpoints is maintained.

This algorithm is far simpler (though far more specialized) than adaptive
mesh refinement.  The code was about 200 lines of Fortran.

Although this algorithm is well suited for a study of the critical solution,
it is not so well suited for a study of the scaling behavior of the mass
of the  black hole.  In treating the critical solution a great deal of
accuracy was gained by choosing the outermost gridpoint to be the null geodesic
that just barely hits the singularity.  However, in treating the formation of
a black hole the outermost gridpoint must be chosen sufficiently far that the
corresponding geodesic does not hit the origin before the formation of a
marginally outer trapped surface.  When treating a range of
$ p $
the outermost gridpoint must satisfy this condition for every spacetime in the
one parameter family.  But then for
$ p $
sufficiently close to
$ p^* $
the spatial resolution will not be good enough to resolve the spatial
structure of the solution.  Therefore only a limitted range of
$ p $
can be treated using this algorithm.

\section {Results}

All runs were done with 300 spatial gridpoints.  The code was developed and
run on a NeXT workstation.  Then for higher accuracy it was run on a Cray Y-MP.
The initial data for the scalar field
$ \Phi $
on the
$ u = 0 $
surface was
\be
\Phi ( r ) = {\phi _0} \, {r^2} \, \exp \left [ - \; {{\left ( {{r \,
- \, {r_0}} \over \sigma } \right ) }^2} \right ]
\ee
where
$ {\phi _0} , \, {r_0} $
and
$ \sigma $
are constants.  Thus the initial value of
$ \Phi $
has essentially a gaussian profile of width
$ \sigma $
centered about a spherical shell of radius
$ r_0 $.
The values of
$ \sigma $
and
$ r_0 $
are fixed, while the amplitude
$ \phi _0 $
is the parameter
$ p $.

The code was first run to find the critical value of the parameter
$ p $.
This was done by a binary search: a parameter that led to the formation
of a black hole
was higher than the critical parameter.  One that did not lead to black hole
formation was smaller than the critical value.  The next estimate of the
critical parameter was chosen halfway between one known to be too high and
one known to be too low.  This process was iterated until the critical
value of the parameter was found to the needed accuracy.  At the same time the
value of the outermost gridpoint was chosen as described in the
previous section.

The code was then run to examine the scaling behavior of the black hole mass.
The outermost gridpoint was chosen large enough to accomodate a range of
the parameter
$ p $
to good accuracy.  The code was then run for several values of
$ p $
in this range.  For each value of
$ p $
the code was run until a marginally outer trapped surface formed.  The mass
$ M $
of the black hole was then calculated.  In figure 1
$ \ln M $
is plotted as a function of
$ \ln \left | p - {p^*} \right | $.

\figloc1{ }

The points are well fit by a straight line whose slope is
$ \approx 0.38 $.
Thus the mass behaves like
$ M \propto {{(p - {p^*} )}^\gamma} $
where
$ \gamma \approx 0.38 $.
This is consistent with the results of references\cite{Chop,GPP}.

Next the code was run to treat the critical solution.  The parameter
$ p $
was set to
$ p^* $
and the outermost gridpoint was set to its optimum value.  Let
$ u^* $
be the value of
$ u $
at which the singularity forms.  Define
$ T $
and
$ R $
by
$ T \equiv - \,  \ln \left ( {u^*} -  u  \right ) $
and
$ R \equiv r {e^T} $.
Then the periodic self-similar property of the critical solution is that
$ h ( R , T ) $
is a periodic function of
$ T $.

\figloc2{ }

In figure 2 the quantity
$ h $
is plotted as a function of
$ R $
and
$ T $.
After a certain amount of evolution the scalar field settles down to a
behavior that seems to be periodic in
$ T $.
To examine this apparent periodicity more carefully we pick an
identifiable set of times: those times at which the maximum of
$ h $
occurs at
$ r = 0 $.
The corresponding values of
$ T $
are
$ {T_1} \approx 2.58 \, , \; {T_2} \approx 6.02 \, , \; {T_3} \approx 9.47 $
and
$ {T_4} \approx 12.95 $.
Note that these times are equally spaced in
$ T $
with a spacing
$ \Delta T = 3.45 $
in agreement with the result found by Choptuik.  We then plot
$ h ( R ) $
at each of these times to see whether this function is the same at each time.

\figloc3{ }

Figure 3 shows
$ h ( R , {T_1} ) $.

\figloc4{ }

Figure 4 shows
$ h ( R , {T_1} ) $
and
$ h ( R , {T_2} ) $
Note that the functions agree.  I emphasize that this figure has two
{\it different } functions
$ h ( R , {T_1} ) $
and
$ h ( R , {T_2} ) $
plotted on the same graph.  The agreement is so good that one cannot tell that
the two functions differ.

\figloc5{ }

Figure 5 shows
$ h ( R , {T_1} ) \, , \;  h ( R , {T_2} ) $
and
$  h ( R , {T_3} ) $
plotted together on the same graph.  Again the agreement is so good
that one cannot tell that the functions differ.

\figloc6{ }

Figure 6 shows
$ h ( R , {T_1} ) \, , \;  h ( R , {T_2} ) \, , \;  h ( R , {T_3} ) $
and
$  h ( R , {T_4} ) $
plotted together on the same graph.  Here the disagreement between the
functions is barely visible.

Thus it is clear that after some initial evolution the scalar field of the
critical solution settles down to an evolution that is periodic in
$ T $.
Therefore we have confirmed, using a completely different algorithm from
that of reference\cite{Chop}, that
the critical solution spacetime has periodic self-similarity.

\section*{Acknowledgments}
It is a pleasure to thank Beverly Berger, Chuck Evans and G. Comer Duncan
for helpful discussions.  I thank the Institute for Theoretical Physics,
University of California, Santa Barbara, the Aspen Center for Physics
and the Astronomy Department of the
University of Michigan for hospitality.  Computations were performed using
the facilities of the National Center for Supercomputing Applications at
The University of Illinois.  This work was supported in part by
National Science Foundation Grant PHY94-08439 and Research Corporation
Grant C-3703 to Oakland University.

\begin{figure}
\caption{Scaling of the black hole mass.  $\ln M $ is plotted vs.
$\ln \left ( p - {p^*} \right )$.
The result is a straight line whose slope is the
critical exponent $\gamma \approx 0.38 $.}
\label{fig1}
\end{figure}

\begin{figure}
\caption{Behavior of $ h $ as a function of the logarithmic coordinates
$ R $ and $ T $.  After some initial evolution $ h $
is a periodic function of $ T $.  This demonstrates the periodic
self-similarity of the spacetime.}
\label{fig2}
\end{figure}

\begin{figure}
\caption{$ h ( R ) $ is plotted at $ T_1 $}
\label{fig3}
\end{figure}

\begin{figure}
\caption{$ h ( R ) $ is plotted at two {\it different} times $ T_1 $ and
$ T_2 $ on the {\it same} graph.  Note that the two functions
agree completely.}
\label{fig4}
\end{figure}

\begin{figure}
\caption{$ h ( R ) $ is plotted at three {\it different} times
$ {T_1} , \, {T_2} $ and $ T_3 $ on the {\it same} graph.  Note that
the three functions agree completely.}
\label{fig5}
\end{figure}

\begin{figure}
\caption{$ h ( R ) $ is plotted at four {\it different} times
$ {T_1} , \, {T_2} , \, {T_3} $ and $ T_4 $ on the {\it same} graph.
One can barely see the disagreement among these functions.}
\label{fig6}
\end{figure}

\end{document}